\documentclass[aps,amsfonts,superscriptaddress,twocolumn]{revtex4-2}

\usepackage[pdftex]{graphicx}
\usepackage{amssymb}
\usepackage{amsmath}
\usepackage{physics}
\usepackage{xspace}
\usepackage{color}
\usepackage{hyperref}
\usepackage{mathtools}

\newcommand\hatT{\mathcal T}

\begin{document}

\title{Probabilistic imaginary-time evolution 
in state-vector-based and shot-based simulations and on quantum devices}

\author{Satoshi Ejima}
\email{satoshi.ejima@dlr.de}
\affiliation{Institute of Software Technology, German Aerospace Center (DLR), 22529 Hamburg, Germany}
\affiliation{Computational Condensed Matter Physics Laboratory, RIKEN Pioneering Research Institute (PRI), Saitama 351-0198, Japan}

\author{Kazuhiro Seki}
\email{kazuhiro.seki@riken.jp}
\affiliation{Quantum Computational Science Research Team, RIKEN Center for Quantum Computing (RQC), Saitama 351-0198, Japan}

\author{Benedikt Fauseweh}
\email{Benedikt.Fauseweh@dlr.de}
\affiliation{Institute of Software Technology, German Aerospace Center (DLR), 51147 Cologne, Germany}
\affiliation{Department of Physics, TU Dortmund University, Otto-Hahn-Str.~4, 44227 Dortmund, Germany}

\author{Seiji Yunoki}
\email{yunoki@riken.jp}
\affiliation{Quantum Computational Science Research Team, RIKEN Center for Quantum Computing (RQC), Saitama 351-0198, Japan}
\affiliation{Computational Quantum Matter Research Team, RIKEN Center for Emergent Matter Science (CEMS), Wako, Saitama 351-0198, Japan}
\affiliation{Computational Materials Science Research Team,
RIKEN Center for Computational Science (R-CCS), Kobe, Hyogo 650-0047, Japan}
\affiliation{Computational Condensed Matter Physics Laboratory,
RIKEN Pioneering Research Institute (PRI), Saitama 351-0198, Japan}

\date{\today}

\begin{abstract}
 Imaginary-time evolution, an important technique in tensor network and quantum Monte Carlo algorithms on classical computers, has recently been adapted to  quantum computing. In this study, we focus on probabilistic imaginary-time evolution (PITE) algorithm and derive its formulation in the context of state-vector-based simulations, where quantum state vectors are directly used to compute observables without statistical errors. We compare the results with those of shot-based simulations, which estimate observables through repeated projective measurements. Applying the PITE algorithm to the Heisenberg chain, we investigate  optimal initial conditions for convergence. We further demonstrate the method on the transverse-field Ising model using a state-of-the-art trapped-ion quantum device. Finally, we explore the potential of error mitigation in this framework, highlighting practical considerations for near-term digital quantum simulations. 
\end{abstract}

\maketitle

\section{Introduction}

As Feynman predicted in 1982~\cite{feynman1982}, quantum computers can naturally encode quantum many-body states, making them ideally suited for simulating  quantum systems. The system Hamiltonian $\hat{\cal H}$ can be mapped onto a 
qubit-based Hamiltonian to construct a quantum circuit, enabling real-time evolution (RTE) via the corresponding unitary dynamics.  By applying  Trotter decomposition, the real-time ($t$) propagator $e^{-\mathrm{i}\hat{\cal H}t}$ can be decomposed into a sequence of single- and two-qubit gates, owing to its unitarity \cite{Lloyd1996}. 
RTE on digital quantum computers allows for the exploration of various quantum 
phenomena, including statistical mechanical properties of quantum many-body states at equilibrium~\cite{Lu2021,Summer2024,Hemery2024,Seki2024} and  non-equilibrium phenomena~\cite{Smith2019,Fauseweh2021}. For example, it has 
enabled the study of discrete time crystals in periodically driven (Floquet) systems, in both one~\cite{ippoliti2021,frey2022,mi2022,camacho2024} and two spatial dimensions~\cite{shinjo2024}.

In materials science, simulating physical systems on quantum computers often begins with the accurate determination of their ground states. One of the most widely used methods for this task is the variational quantum eigensolver (VQE)~\cite{Peruzzo2014,mcclean2016,kandala2017,Lively2024}. VQE uses the variational principle to approximate the ground state by optimizing a parameterized quantum circuit to minimize the expectation value of the energy. This hybrid quantum-classical algorithm combines classical optimization techniques with quantum circuit evaluations, making it particularly well suited for near-term quantum computers, especially noisy intermediate-scale quantum (NISQ) devices~\cite{preskill2018,RevModPhys.94.015004,Fauseweh2023,fauseweh2024}.
However, the difficulty of ground-state search increases substantially with system size. A major obstacle is the so-called barren plateau phenomenon, where the optimization landscape becomes exponentially flat as the number of qubits increases~\cite{McClean_2018}. This leads to vanishing gradients, hindering the convergence of classical optimizers and thus limiting the scalability of VQE for large-scale simulations.

A promising alternative is imaginary-time evolution (ITE). By introducing imaginary time $\tau=\mathrm{i}t$ and applying the propagator $e^{-\hat{\cal H}\tau}$ to an initial state $\ket{\psi(0)}$, the evolved state is obtained as $\ket{\psi(\tau)}=\gamma e^{-\hat{\cal H}\tau}\ket{\psi(0)}$, where $\gamma$ is a normalization constant.
In classical computation, ITE has proven useful for various quantum problems, including ground-state search and finite-temperature simulations. Established method such as quantum Monte Carlo~\cite{QMC1987}, time-evolving block decimation~\cite{TEBD}, and density-matrix renormalization group~\cite{PhysRevB.72.220401} techniques have successfully employed ITE in these contexts. Unlike  RTE, however, implementing ITE on a quantum computer poses a challenge: the propagator $e^{-\hat{\cal H}\tau}$ is non-unitary and thus cannot be directly decomposed into a sequence of quantum gates using conventional Trotterization techniques. This limitation necessitates alternative strategies for realizing ITE on quantum hardware.  

Several ITE algorithms tailored for quantum computation have been proposed, including variational ITE~\cite{mcardle2019,PhysRevA.99.062304,Yuan2019}, quantum ITE~\cite{Motta2019,YeterAydeniz2020,sun2021}, and probabilistic ITE (PITE)~\cite{lin2021,liu2021,PITE,PITE2,Leadbeater_2024}. These methods have been mainly demonstrated on small-scale systems, such as simple molecular systems (e.g., H$_2$ and LiH) and quantum spin chains with system sizes $L<10$. 
To advance digital quantum simulations beyond these benchmarks, it is essential to explore the applicability of ITE to larger-scale systems. This is particularly timely given recent progress in hardware, such as trapped-ion quantum processor, which feature high fidelities, all-to-all connectivity, and device size exceeding 50 qubits. 
Furthermore, the choice of initial state as well as the initial conditions---an important yet underexplored aspect---play a crucial in the efficiency and accuracy of ITE and are highly dependent on the target system, system size, and Trotter time step $\Delta\tau$.

In this study, we apply the PITE algorithm, using only a single  ancilla qubit~\cite{PITE,PITE2}, to standard spin chains with system sizes up to $L=20$. Our aim is to identify potential bottlenecks in scaling the method to larger systems. To this end, we derive a  state-vector-based formulation of the PITE algorithm, which is particularly well suited for simulating larger systems on classical computers. This formulation also allows us to systematically determine optimal initial parameters, leading to success probabilities approaching unity. 
The PITE algorithm involves controlled RTE operations, which entangle the ancilla qubit with all system qubits. 
Hence, implementing the algorithm on NISQ devices remains challenging, especially in obtaining reliable outputs, even after applying error mitigation techniques. 
To evaluate its practical feasibility, we perform modified PITE simulations with up to 17 qubits on a trapped-ion quantum computer.  These experiments utilize the optimized initial parameters obtained from our state-vector-based analysis and employ multiple ancillary qubits---equal to the number of imaginary-time steps---to reduce the number of required reset operations.

The rest of this paper is organized as follows. In Sec.~\ref{sec:pite}, we briefly review the PITE algorithm. In Sec.~III, we derive a state-vector-based formulation of the PITE algorithm. 
Section IV presents numerical results obtained from both state-vector-based and shot-based simulations, demonstrating perfect agreement between the two approaches, as expected. Based on these results, we investigate the optimization of the initial parameters to maximize success probabilities in representative spin systems. In Sec.~V, we present experimental results obtained on a trapped-ion quantum computer, employing the optimized initial parameters and discussing potential error mitigation strategies. We conclude in Sec~VI with a summary and outlook for future research. System-size dependence is examined in the \hyperref[appen:size]{Appendix}.

\section{PITE algorithm}
\label{sec:pite}

\begin{figure}[tb]
    \centering
    \includegraphics[height=1.5cm]{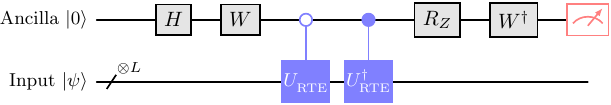}
    \caption{Quantum circuit of the approximate PITE algorithm for a single imaginary-time step~\cite{PITE}. $H$ denotes the Hadamard gate and $R_Z\equiv R_Z(-2\theta_0)$ represents a single-qubit rotation about the $Z$ axis.}
    \label{fig:circuit_pite}
\end{figure}

In this study, we follow the PITE algorithm as formulated in Refs.~\cite{PITE} and \cite{PITE2}. 
Let us consider a spin-1/2 system with $L$ sites, described by a Hamiltonian $\hat{\cal H}$. Starting from an initial state $|\psi_{\rm ini}\rangle$, the goal is to implement ITE for a time increment $\Delta \tau$, i.e., to apply the operator 
$e^{-\hat{\cal H}\Delta\tau}$. 
To this end, we define a nonunitary Hermitian operator 
$\hat{\mathcal T}\equiv\gamma e^{-\hat{\cal H}\Delta\tau}$,
where $\gamma$ is a tunable real parameter satisfying 
$0<\gamma < 1$ and $\gamma\neq 1/\sqrt{2}$.
We embed the nonunitary operator $\hat{\mathcal T}$ into a unitary matrix of the form  
\begin{align}
    \mathcal{U}_{\hatT}
    \equiv
    \begin{pmatrix}
        \hatT & \sqrt{1-\hatT^2} \\
        \sqrt{1-\hatT^2} & -\hatT
    \end{pmatrix}\,.
\end{align}
Introducing an ancillary qubit initialized in the state $\ket{0}$, we obtain 
\begin{align}
 \mathcal{U}_{\hatT}\ket{\psi}\otimes\ket{0}
  =\hatT\ket{\psi}\otimes\ket{0}
   +\sqrt{1-\hatT^2}\ket{\psi}\otimes\ket{1}\,.
\end{align}
Thus, upon measuring the ancilla qubit  in the $\ket{0}$ state, 
which occurs with the probability $\mathbb{P}_0=\bra{\psi}\hatT^2\ket{\psi}$, the post-measurement state of the system is given (up to normalization) by 
\begin{align}
    \ket{\Psi(\tau)}=\frac{1}{\sqrt{\mathbb{P}_0}}
    \hatT\ket{\psi}\,.
\end{align}
The quantum circuit representing the unitary $\mathcal{U}_{\hatT}$ contains $e^{\pm\mathbf{i}\kappa\Theta}$, where  $\kappa=\mathrm{sgn}(\gamma-1/\sqrt{2})$ and
$\Theta\equiv \arccos\left[(\hatT+\sqrt{1-\hatT^2})/\sqrt{2}\right]$.

Since directly implementing $e^{\pm\mathbf{i}\kappa\Theta}$ in quantum circuits is not feasible, we instead approximate $\Theta$ using a first-order Taylor expansion in $\Delta\tau$:  
\begin{align}
    \kappa\Theta=\theta_0-\hat{\cal H}s_1\Delta\tau+\mathcal{O}(\Delta\tau^2)\,,
\end{align}
where $\theta_0=\kappa\arccos[(\gamma+\sqrt{1-\gamma^2})/\sqrt{2}]$ and 
$s_1=\gamma/\sqrt{1-\gamma^2}$. 
This approximation allows us to express $e^{\pm\mathbf{i}\kappa\Theta}$ in terms of RTE operators 
$\hat{U}_{\rm RTE}(\Delta t)\equiv e^{-\mathrm{i}\hat{\cal H}\Delta t}$ as follows: 
\begin{align}
    e^{\mathbf{i}\kappa\Theta}&\otimes\ket{0}\bra{0}
    +e^{-\mathbf{i}\kappa\Theta}\otimes\ket{1}\bra{1}
    \notag\\
    &=\left(I_{2^L}\otimes R_z(-2\theta_0)\right)
      \left(
       \hat{U}_{\rm RTE}(s_1\Delta\tau)\otimes\ket{0}\bra{0}
       \right.
       \notag\\
      &+\left.\hat{U}_{\rm RTE}^\dagger(s_1\Delta\tau)\otimes\ket{1}\bra{1}\right)\,.
\end{align}
Using this decomposition, the PITE quantum circuit can be constructed as illustrated in Fig.~\ref{fig:circuit_pite}, where the single-qubit gate $W$ is defined as
\begin{align}
    W\equiv
    \frac{1}{\sqrt{2}}
    \begin{pmatrix}
        1 & -\mathrm{i} \\
        1 & \mathrm{i}
    \end{pmatrix}\,.
\end{align}

Obviously, the accuracy of the time evolution deteriorates with increasing $\Delta\tau$, necessitating the use of sufficiently small time steps. 
Furthermore, the algorithm requires the projective measurement of the ancillary qubit after each step and proceeds only when the $\ket{0}$ outcome (success state) is obtained. Repeating this procedure filters out all excited-state components of the target system, thereby projecting onto the ground state. 
 To minimize shot loss in shot-based simulations or on actual quantum devices, it is essential to maximize the success probability at each step. This can be achieved by carefully choosing the optimal initial parameters, $\Delta t$ and $\gamma$, such that the probability of measuring the ancilla in the success state approaches unity. 
 In the following section, we derive the PITE formulation based on state-vector simulations, which enables efficient evaluation of optimal parameter sets when the system size is small enough to fit into the classical memory.

\section{State-vector simulation method} 

In general, it is highly advantageous to execute quantum algorithms using state-vector simulators  whenever feasible, as they are significantly faster than shot-based simulators, even in the absence of noise. This advantage arises from the fact that state-vector simulators provide exact computations of observables without statistical errors, whereas shot-based simulators emulate the behavior of quantum devices through repeated sampling. In this section, we begin by revisiting the well-known Hadamard test, which is widely used to estimate the expectation value of a unitary operator $\hat{U}$ on quantum hardware. This serves as a simple illustrative example due to its structural similarity to the state-vector-based simulations of the PITE algorithm, which we  discuss in the latter part of this section.

\subsection{Hadamard test}

\begin{figure}[tb]
    \centering
    \includegraphics[height=1.5cm]{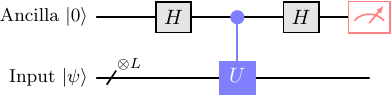}
    \caption{Quantum circuit for the Hadamard test.}
    \label{fig:Hadamard-test}
\end{figure}

Before discussing the state-vector simulation of the PITE algorithm, 
we briefly revisit the Hadamard test as a simple illustrative example. 
The quantum circuit for this test is shown in Fig.~\ref{fig:Hadamard-test}. 
The first qubit is initialized to $\ket{0}$, while the remaining qubits are initialized to 
the target state $\ket{\psi}$. A Hadamard gate is first applied to the ancillary qubit, followed by 
a controlled unitary operator $\hat{U}$, which applies $\hat U$ only if the ancillary qubit is 
the state $\ket{1}$ and acts as the identity otherwise: 
\begin{align}
 \ket{0}\otimes\ket{\psi}
  &\xmapsto{H, \hat{U}}
  \frac{1}{\sqrt{2}}
   \left(
    \ket{0}\otimes\ket{\psi}
    +\ket{1}\otimes \hat{U}\ket{\psi}
   \right)
   \notag\\
  &\xmapsto{H}
  \ket{0}\otimes\frac{1+\hat{U}}{2}\ket{\psi}
  +\ket{1}\otimes\frac{1-\hat{U}}{2}\ket{\psi}
  \,.
\end{align}
By measuring the ancillary qubit, we obtain probabilities $p_0$ and $p_1$ of observing outcomes 
$0$ and $1$, respectively, from which the real part of the expectation value of $\hat U$ can be 
extracted as  
\begin{align}
 p_0=\frac{1}{2}(1+{\rm Re}\bra{\psi}\hat{U}\ket{\psi})\,,
 \ \ \ 
 p_1=\frac{1}{2}(1-{\rm Re}\bra{\psi}\hat{U}\ket{\psi})\,. 
\end{align}
Thus, the real part is obtained via ${\rm Re}\bra{\psi}\hat{U}\ket{\psi}=p_0-p_1$.

In contrast, with a state-vector simulator, the expectation value $\bra{\psi}\hat{U}\ket{\psi}$ can 
be computed directly, as the state $\ket{\psi}$ is explicitly available. This eliminates the need 
for the ancillary qubit and measurement operations, resulting in a significant reduction in 
computational time.

\subsection{State-vector PITE}

In this subsection, we demonstrate that the approximate PITE algorithm (Fig.~\ref{fig:circuit_pite}) 
can be implemented as state-vector-based simulation, analogous to the Hadamard test discussed in the 
previous subsection. Let $\ket{\psi_j}$ denote the quantum state after $j$th imaginary-time step. 
Applying the $(j+1)$th imaginary-time step to $\ket{\psi_{j}}$, the circuit evolves the system as follows: 
\begin{align}
 \ket{0}&\otimes\ket{\psi_j} 
 \xmapsto{H,\, W} 
 \left(
    \frac{1-\mathrm{i}}{2}\ket{0}+\frac{1+\mathrm{i}}{2}\ket{1}
 \right)\otimes\ket{\psi_j} 
 \nonumber\\
 &\xmapsto{\hat{U},\, \hat{U}^\dagger} 
  \ket{0}\otimes\frac{1-\mathrm{i}}{2}\hat{U}_{\rm RTE}\ket{\psi_j}
  +\ket{1}\otimes\frac{1+\mathrm{i}}{2}\hat{U}_{\rm RTE}^\dagger\ket{\psi_j}
  \nonumber\\
 & \xmapsto{R_z,\, W^\dagger} 
 \ket{0} \otimes 
 \frac{\sqrt{2}}{4}\left(
  (1-\mathrm{i}) e^{\mathrm{i}\theta_0}\hat{U}_{\rm RTE}
  +{\rm h.c.}
 \right) \ket{\psi_j}
 \nonumber\\
 &\hspace{1.2cm}+ \ket{1}\otimes
 \frac{\sqrt{2}}{4}\left(
  (1+\mathrm{i}) e^{\mathrm{i}\theta_0}\hat{U}_{\rm RTE}
  +{\rm h.c.}
 \right) \ket{\psi_j}\,.
\end{align}
By projecting onto the success state $\ket{0}$ of the ancillary qubit, we obtain the (unnormalized) 
updated wavefunction after the $(j+1)$th step as  
\begin{align}
 \ket{\psi_{\rm new}} = 
 \frac{1}{2\sqrt{2}}\left(
  (1-\mathrm{i}) e^{\mathrm{i}\theta_0}\hat{U}_{\rm RTE}
 +(1+\mathrm{i}) e^{-\mathrm{i}\theta_0}\hat{U}^\dagger_{\rm RTE} 
 \right) \ket{\psi_{j}}\,.
 \label{eq:SV-PITE}
 \end{align}
The normalized wavefunction is then given by  
$\ket{\psi_{j+1}}=\ket{\psi_{\rm new}}/\sqrt{\mathbb{P}_0^{(j+1)}}$ , 
where the success probability after the $(j+1)$th step is 
$\mathbb{P}_0^{(j+1)}=\bra{\psi_{\rm new}}\ket{\psi_{\rm new}}$.  
It should be reminded that $\mathbb{P}_0^{(j)}$ is a conditional probability that represents the probability of success at $(j+1)$th imaginary-time step under the condition that the first, second, $\dots$, and $j$th imaginary-time steps were successful. 
Finally, the energy of the target system with Hamiltonian $\hat{\cal H}$ can be estimated as 
$\bra{\psi_{j+1}}\hat{\cal H}\ket{\psi_{j+1}}$.

As is evident from Eq.~\eqref{eq:SV-PITE}, 
the state-vector-based PITE does not require the use of 
an ancillary qubit, nor does it involve the mid-circuit measurements or resets at the end or 
beginning of each imaginary-time step. 
Moreover, executing the algorithm in the state-vector framework is equivalent to performing 
a shot-based simulation with an infinite number of shots. This allows simulations with low success 
probabilities to be carried out 
as many time steps as desired, without incurring a loss of statistical precision. 
As a result, the optimal choice of initial parameters $\gamma$ and $\Delta\tau$ 
can be determined most efficiently through state-vector simulations. 
In addition, the cumulative survival rate after $j$ time steps in a shot-based simulation 
or real-device implementation can be estimated as 
$N_{\rm shots}^{(j)}=N_{\rm shots}^{(0)}\prod_{i=1}^{j} \mathbb{P}_0^{(i)}$, 
where $N_{\rm shots}^{(0)}$ is the initial number of shots and $\mathbb{P}_0^{(i)}$ is the 
success probability at the $i$th time step. 

It should be noted, however, that the state-vector simulations become infeasible when 
the size of the Hamiltonian for the target system exceeds the available classical memory capacity. 
The number of qubits manageable with a state-vector simulator on a classical supercomputer will be at most around 48~\cite{DeRaedt2019}.

\section{Results of numerical simulations}

In this section, we present results from both state-vector-based and shot-based simulations, 
conducted using the IBM Qiskit library~\cite{qiskit}.

\subsection{Heisenberg model}

We consider the isotropic spin-1/2 Heisenberg chain of $L$ sites. The Hamiltonian is given by 
\begin{align}
    \hat{\cal H}_{\rm Heisen} 
    = \frac{J}{4}\sum_j\left(
    \hat{X}_j\hat{X}_{j+1}+\hat{Y}_j\hat{Y}_{j+1}+\hat{Z}_j\hat{Z}_{j+1}
    \right)\,,
    \label{eq:Heisen}
\end{align}
where $\hat{X}_j$, $\hat{Y}_j$ and $\hat{Z}_j$ are the Pauli operators acting on site $j$, 
and $J>0$ denotes the antiferromagnetic exchange interaction. 
Periodic boundary conditions (PBC) are imposed such that the index $j+1$ is interpreted as 1 
when $j=L$. Throughout this study, we set $J=1$ to define the energy unit.

\begin{figure}[tb]
    \centering
    \includegraphics[scale=0.8]{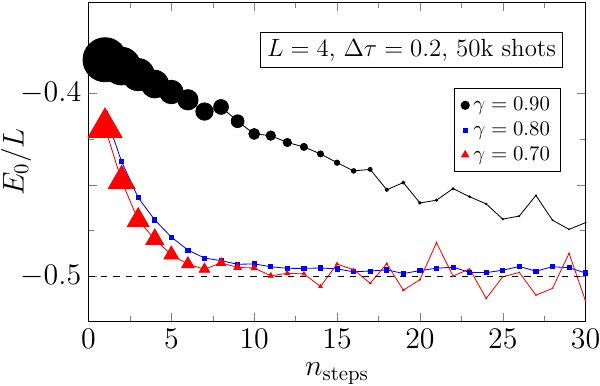}
    \caption{Ground-state energy per site $E_0/L$ vs. the number $n_{\rm step}$ of imaginary-time 
    steps obtained using the PITE algorithm for various 
    values of $\gamma$ in the Heisenberg chain with $L=4$ under PBC. Results are from the 
    noiseless shot-based simulations. The symbol sizes for $\gamma=0.7$ and $0.9$ are proportional 
    to the number of successful outcomes after each imaginary-time step. The dashed line indicates the exact 
    value of $E_0/L$, while the solid lines are guides to the eye.}
    \label{fig:L4-e0-shot}
\end{figure}

Figure~\ref{fig:L4-e0-shot} shows noiseless shot-based PITE results for the Heisenberg 
model with $L=4$. The initial state $\ket{\psi_{\rm ini}}$ is chosen to be the 
singlet state, which yields significantly faster convergence and higher accuracy compared to 
starting from the antiferromagnetic product state $\ket{\uparrow,\downarrow,\uparrow,\downarrow}$. 
For a fixed imaginary-time step $\Delta\tau=0.2$, when $\gamma$ is either too small ($\gamma=0.7$) 
or too 
large ($\gamma=0.9$), the number of successful shots in each imaginary-time step decreases due to low success 
probabilities, as indicated by the size of the symbols in the figure. Starting 
$N_{\rm shots}^{(0)}=50\,000$ shots, only $N_{\rm shots}^{(30)} < 500$ ($1500$) shots remain 
after 30 imaginary-time steps for $\gamma=0.7$ ($0.9$), resulting in unstable estimates beyond 10 (15) steps. 
In contrast, for $\gamma=0.8$, more than 25\,000 shots survive even after 30 steps, 
yielding stable results beyond 20 steps. 
These observations highlight the critical importance of choosing an optimal initial parameter set 
for $\Delta\tau$ and $\gamma$ that maximizes the success probability $\mathbb{P}_0$.

\begin{figure}[tb]
    \centering
    \includegraphics[scale=0.8]{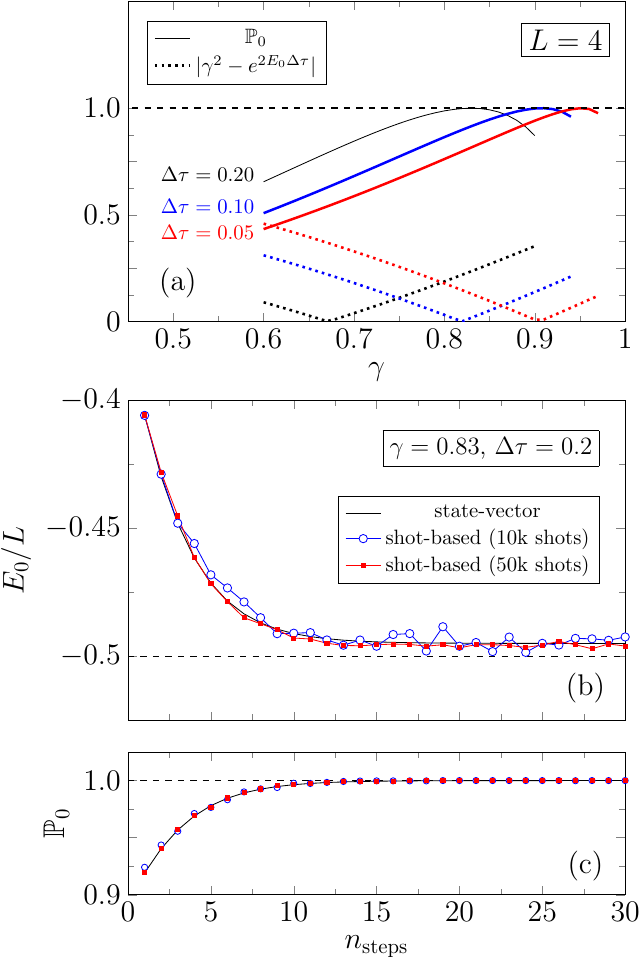}
        \caption{(a) $\gamma$ dependence of the success probability $\mathbb{P}_0^{(100)}$ 
        (solid lines) and $|\gamma^2-e^{2\Delta\tau E_0}|$ (dotted lines) for various 
        imaginary-time steps $\Delta\tau$ after 100 imaginary-time steps in the Heisenberg chain with $L=4$ 
        under PBC, computed using state-vector simulations.
    (b) Estimated ground-state energy for $\Delta\tau=0.2$ and $\gamma=0.83$ using 
    state-vector simulations and shot-based simulations with $N_{\rm shots}^{(0)}=10\,000$ and 50\,000. 
    (c) Corresponding success probabilities $\mathbb{P}_0$ for each imaginary-time step in panel (b).
    }
    \label{fig:sv-shot-pite}
\end{figure}

State-vector-based PITE simulations offer a valuable opportunity to optimize the initial parameters, 
as long as the system size of the target Hamiltonian fits within the main memory of the classical 
computer used. 
Figure~\ref{fig:sv-shot-pite}(a) shows the $\gamma$ dependence of the success probability 
$\mathbb{P}_0$ for different values of $\Delta\tau$ (solid lines). Clearly, $\mathbb{P}_0(\gamma)$ 
exhibits a peak structure; for instance, at $\Delta\tau = 0.2$, the peak appears near 
$\gamma_{\rm max}\approx0.83$, where $\mathbb{P}_0\simeq 1$. 
A natural question arises: Does this $\gamma_{\rm max}$ yield a final state sufficiently close to 
the ground state? To address this, let us first consider an idealized situation. Suppose that 
after several imaginary-time steps, the system reaches the ground state $\ket{\psi_0}$, so that the success 
probability becomes 
$\mathbb{P}_0=\bra{\Psi}\hat{\cal T}^2\ket{\Psi}\simeq \gamma^2e^{-2\Delta\tau E_0}$. 
Imposing the desired condition $\mathbb{P}_0\simeq 1$ leads to the optimal choice 
$\gamma_{\rm opt}=e^{E_0\Delta\tau}$. 
However, the observed peak position $\gamma_{\rm max}$ in $\mathbb{P}_0(\gamma)$ does not 
coincide with this $\gamma_{\rm opt}$, as shown by the deviation 
$\epsilon(\gamma)=|\gamma^2-e^{2E_0\Delta\tau}|$ [dotted lines in Fig.~\ref{fig:sv-shot-pite}(a)]. 
For example, at $\Delta\tau=0.2$, $\epsilon(\gamma)$ becomes minimal at 
$\gamma\simeq0.67\, (=\gamma_{\rm opt})$, while $\gamma_{\rm max} > \gamma_{\rm opt}$. 
To understand this discrepancy, let us consider the correction to $\ket{\Psi}$ and 
define a state $\ket{\Psi^\prime}$ with  
$\bra{\Psi^\prime}\hat{\cal T}^2\ket{\Psi^\prime}=\gamma^2e^{-2E_1\Delta\tau}$ and 
$E_1=E_0+\Delta E\geq E_0$. Because of $\gamma_{\rm max}^2e^{-2E_1\Delta\tau}\simeq 1$, 
\begin{align}
    \frac{\gamma_{\rm max}}{\gamma_{\rm opt}}
        =\frac{e^{E_1\Delta\tau}}{e^{E_0\Delta\tau}}=e^{\Delta E\Delta\tau}\geq 1\,.
\end{align}
This shows that $\gamma_{\rm max}$, which satisfies $\mathbb{P}_0(\gamma_{\rm max})\simeq 1$, 
is always strictly larger than $\gamma_{\rm opt}$ corresponding to the ideal ground state. 
Moreover, as $\Delta\tau$ decreases, $\gamma_{\rm max}$ gradually approaches $\gamma_{\rm opt}$, 
which is consistent with the trend observed in Fig.~\ref{fig:sv-shot-pite}(a).

Figure~\ref{fig:sv-shot-pite}(b) shows the estimation of the ground-state energy per site 
($E_0/L$) at $\gamma=0.83\, (=\gamma_{\rm max})$ for $\Delta\tau=0.2$, obtained from both 
state-vector and shot-based simulations. 
After 30 imaginary-time steps, the results from the state-vector simulation, $E^{\rm SV}_0/L\simeq 0.495$ 
[solid line in Fig.~\ref{fig:sv-shot-pite}(b)], exhibit excellent agreement with the exact 
value $E^{\rm ex}_0/L=-0.5$, obtained by the exact-diagonalization (ED) method. 
The shot-based results (symbols) also show good agreement with the state-vector simulation, 
especially as the number of shots increases. This is expected, since the state-vector simulation is effectively equivalent to the shot-based simulation in the limit of an infinite number of shots. 
The corresponding success probabilities $\mathbb{P}_0$ for each imaginary-time step are shown in 
Fig.~\ref{fig:sv-shot-pite}(c). For the finely tuned parameters $\gamma=0.83$ and $\Delta\tau=0.2$, the success probability $\mathbb{P}_0$ approaches unity after approximately 10 imaginary-time steps in both the state-vector and shot-based simulations.

\begin{figure}[tb]
    \centering
    \includegraphics[scale=0.75]{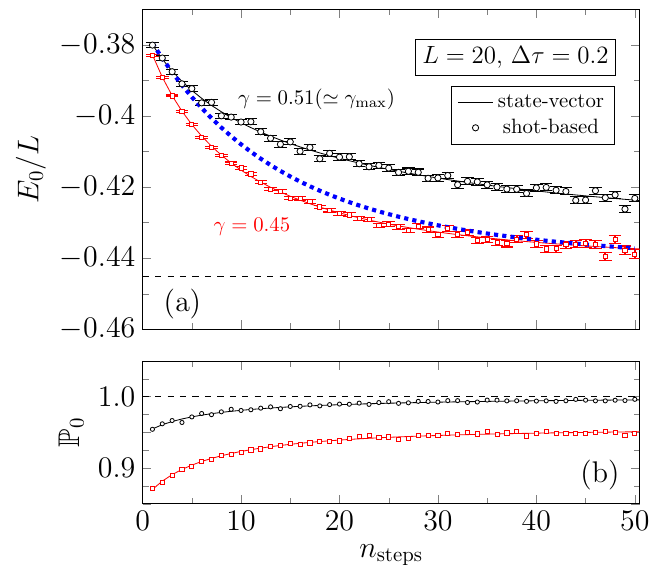}
    \caption{
    (a) Ground-state energy per site $E_0/L$ and (b) success probability $\mathbb{P}_0$ 
    as functions of the number $n_{\rm step}$ of imaginary-time steps, 
    obtained using the PITE algorithm with $\gamma=0.45$ (squares) and $0.51$ (circles) for the 
    Heisenberg chain with $L=20$ under PBC. 
    Solid lines represent the results from  the state-vector simulations, while 
    the dotted line in panel (a) shows the classical ITE simulation for comparison. 
    Error bars in panel (a) indicate the statistical uncertainty $\sigma_E$ due to sampling, 
    estimated from the standard deviations of the individual energy components $E_{XX}$, $E_{YY}$, 
    and $E_{ZZ}$ as $\sigma_E=\sqrt{\sigma_{XX}^2+\sigma_{YY}^2+\sigma_{ZZ}^2}$\,. 
    }
    \label{fig:L20-Heisenberg}
\end{figure}

Figure~\ref{fig:L20-Heisenberg} presents the PITE results for the larger system size of $L=20$. 
Based on the state-vector simulations (not shown), the optimal initial parameter set is estimated 
as $\gamma_{\rm max}\simeq 0.51$ for the fixed imaginary-time step $\Delta\tau=0.2$, 
consistent with the trend observed in the $L=4$ case [see Fig.~\ref{fig:sv-shot-pite}(a)]. 
As the number of imaginary-time steps increases, the estimated ground-state energy per site $E_0/L$ 
gradually approaches the exact value $E_0^{\rm ex}/L\simeq-0.445$. However, the convergence 
is slower and the deviation from the exact value is larger than that in the $L=4$ case, 
requiring more imaginary-time steps to reach similar accuracy. 
This behavior can be understood as follows. Let the quantum state of the target system be 
expressed as $\ket{\psi}=\sum_{n=0} c_n\ket{E_n}$, where $\hat{\cal H}\ket{E_n}=E_n\ket{E_n}$ 
with $E_0<E_1\leq E_2\leq \cdots$ and $c_0\neq0$. 
After applying the ITE, the state becomes 
\begin{align}
    e^{-\tau\hat{\cal H}}\ket{\psi}=c_0e^{-\tau E_0}\left\{
    \ket{E_0}+\frac{c_1}{c_0}e^{-\tau (E_1-E_0)}\ket{E_1}+\dots
    \right\}\,,
\end{align}
which shows that $\tau\gg 1/(E_1-E_0)$ is required to sufficiently suppress the contributions 
from excited state. Therefore, for larger systems, where the energy gap $E_1-E_0$ generally decreases 
with increasing $L$, a greater number of imaginary-time steps 
(i.e., $n_{\rm steps}\cdot\Delta\tau\gg 1/(E_1-E_0)$) is needed to reach the ground state 
with high accuracy.

Alternatively, instead of using the optimal parameter $\gamma_{\rm max}(\simeq 0.51)$, one can 
start with a slightly smaller value, such as $\gamma=0.45$. As shown in 
Fig.~\ref{fig:L20-Heisenberg}(a), this choice leads to a faster approach to the ground-state energy 
in shorter imaginary-time steps. However, this comes at the cost of a lower success probability, 
resulting in a reduced number of surviving shots, as shown in Fig.~\ref{fig:L20-Heisenberg}(b). 
More specifically, starting with $N_{\rm shots}^{(0)}=100\,000$ shots, only about 
$N_{\rm shots}^{(50)}<4000$ survive after 50 imaginary-time steps when $\gamma=0.50$. In contrast, 
for $\gamma=0.51$, $N_{\rm shots}^{(50)}<6000$ survive even when $N_{\rm shots}^{(0)}=10\,000$. 
Thus, while smaller $\gamma$ can accelerate convergence in imaginary time, it significantly increases 
the sampling cost in shot-based simulations.

In Fig. \ref{fig:L20-Heisenberg}(a), we also show the numerical results of classical ITE simulations 
performed under the same conditions---starting from the singlet state with $\Delta\tau=0.2$. 
These results show reasonable agreement with those obtained from the PITE simulations. 
An improved implementation could involve adaptively tuning $\Delta\tau$ 
and $\gamma$, for example, by employing optimized imaginary-time steps 
that minimize the energy expectation value for a fixed number of imaginary-time steps~\cite{Yanagisawa_1998,Beach2019,PhysRevResearch.3.013004}.
However, in the case of PITE, a lower energy expectation value does not necessarily correspond 
to a higher success probability, as evidenced in Fig.~\ref{fig:L20-Heisenberg}.

Finally, we discuss the required number of shots to estimate an observable $\hat O$ within a certain statistical uncertainty $\epsilon$, i.e., $\delta O\le \epsilon$.
The effective number of shots at $j$th imaginary-time step is given by  $N_{\rm shots}^{(j)}
=N_{\rm shots}^{(0)}\prod_{i=1}^{j} \mathbb{P}_0^{(i)}
=:N_{\rm shots}^{(0)}\mathbb{P}(j)$.
Assuming that $\delta O \simeq 1/(N_{\rm shots}^{(j)})^{1/2}$,
the number of shots required can be estimated as $N_{\rm shots}^{(j)}\gtrsim 1/({\epsilon}^2\mathbb{P}(j))$ in the noiseless case.

\subsection{Transverse-field Ising model}

Even on state-of-the-art trapped-ion quantum computers, the Heisenberg chain discussed above 
remains too complex for practical implementation of the PITE algorithm on real hardware. 
Therefore, we turn to a simpler model, namely, the one-dimensional transverse-field Ising model 
(TFIM), whose Hamiltonian is given by 
\begin{align}
    \hat{\cal{H}}_{\rm TFIM}=-\sum_{j=1}^L \hat{Z}_j\hat{Z}_{j+1}
    -\sum_{j=1}^L\hat{X}_j\, ,
    \label{eq:HTFIM}
\end{align}
where $\hat{X}_j$ and $\hat{Z}_j$ are the Pauli operators acting on site $j$. As the initial state, 
we use a superposition state in the $XY$ plane by applying a Hadamard gate to each qubit 
initialized in the $\ket{0}$ state.

\begin{figure}[tb]
    \centering
    \includegraphics[scale=0.8]{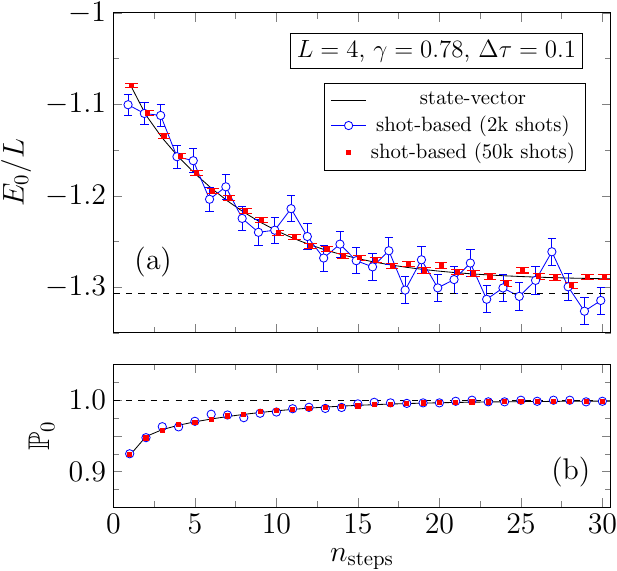}
        \caption{(a) Ground-state energy per site $E_0/L$ and (b) success probability $\mathbb{P}_0$ 
        as functions of the number $n_{\rm step}$ of imaginary-time steps, 
        obtained using the PITE algorithm with $\gamma=0.78$ for the TFIM with $L=4$ under PBC. 
        Solid lines represent the results from the state-vector simulations, while symbols 
        denote those from the shot-based simulations with $N_{\rm shots}^{(0)}=2000$ shots 
        (open blue circles) and $50\,000$ shots (solid red squares). 
        Error bars in panel (a) indicate the statistical uncertainty $\sigma_E$ due to sampling, 
        estimated from the standard deviations of the individual energy components as  
        $\sigma_E=\sqrt{\sigma_X^2+\sigma_{ZZ}^2}$.
    }
    \label{fig:TFIM-simulator}
\end{figure}

Figure~\ref{fig:TFIM-simulator} shows the PITE results for the ground-state energy of the TFIM 
with $L=4$ under PBC, obtained using both state-vector and shot-based simulations. For the fixed 
imaginary-time step $\Delta\tau=0.1$, we identify the optimal parameter 
$\gamma_{\rm max}\simeq 0.78$, as in the case of the Heisenberg chain discussed in 
Fig.~\ref{fig:sv-shot-pite}(a).
As the imaginary-time steps are iterated, the energy per site $E_0/L$ steadily converges toward the exact 
ground-state value $E_0^{\rm ex}/L\simeq -1.31$ on both simulations. Here, we include results 
obtained using only 2000 shots (blue symbols), to reflect the limited number of measurements 
available on real devices, as will be discussed later. As expected, the results with 2000 shots 
exhibit larger statistical fluctuations compared to those using $50\,000$ shots (red squares) or 
the state-vector simulation (black solid line). 
In all cases, the success probability $\mathbb{P}_0$ approaches unity after approximately 15 imaginary-time 
steps [see Fig.~\ref{fig:TFIM-simulator}(b)], as intended by fine choosing the parameter 
$\gamma=0.78\simeq \gamma_{\rm max}$ for $\Delta\tau=0.1$.

\section{Results on a quantum device}

In this section, we demonstrate the implementation of the PITE algorithm for the TFIM with $L=4$ 
under PBC, using a trapped-ion quantum device H1-1 provided by Quantinuum. 
Our primary objective here is to showcase the feasibility and practical behavior of the PITE 
algorithm on actual quantum hardware, within the limitations of currently available quantum 
resources. To this end, we fix the imaginary-time step to $\Delta \tau=0.1$ and use the optimal 
parameter $\gamma=0.78$, as determined in the previous section.

\subsection{Hardware specifications}

The experiments were conducted between early November and early December 2024. 
The specifications of the Quantinuum H1-1 system at the time of experiments are 
summarized below~\cite{HSeries}. 
The H1-1 system comprises 20 qubits and features all-to-all qubit connectivity.  
The average single-qubit and two-qubit gate infidelities were approximately $2\times 10^{-5} $ 
and $1 \times 10^{-3}$, respectively. The average state preparation and measurement (SPAM) error 
rate was $3 \times 10^{-3}$. 
The native two-qubit gate is the $\mathrm{ZZPhase}(\alpha)$ gate, defined as 
$\mathrm{ZZPhase}(\alpha):=\mathrm{e}^{-\frac{1}{2}{\mathrm{i}\pi\alpha} \hat{Z}_i \hat{Z}_j}$, 
which can be applied between any pair of qubits $(i,j)$ with an arbitrary rotation angle $\alpha$.  
For additional technical details, we refer the reader to Ref.~\cite{HSeries}.

\subsection{Experimental setup}

We perform the PITE algorithm up to $n_{\rm steps} = 13$ steps. 
Although the algorithm can be implemented with only $L+1$ qubits by employing mid-circuit 
measurement and reset (MCMR), we instead use $L+n_{\rm steps}$ qubits to avoid MCMR and 
simplify the quantum operations. 
As an initial state $|\psi\rangle$, we use the state polarized along the Pauli-$X$ axis, i.e., 
$|\psi\rangle = |+\rangle :=(\hat{H}|0\rangle)^{\otimes L}$.

The circuits are compiled with the pytket compiler~\cite{Sivarajah_2020}. 
The number of native two-qubit ZZPhase gates in the complied circuit for the $n_{\rm steps}$th 
imaginary-time step is found to be $4 + 27 n_{\rm steps}$ for $n_{\rm steps}\geqslant 1$, 
and $0$ for $n_{\rm steps}=0$. Thus, at $n_{\rm steps}=13$, our circuits contain up to $355$ 
ZZPhase gates, and we utilize a maximum of 17 qubits of the H1-1 system. 

We evaluate the energy expectation value of the TFIM. 
To this end, we measure the $n_{\rm steps}$ ancilla qubits in $Z$ basis and the $L$ system qubits 
in both $Z$ and $X$ bases. 
We postselect the successful application of the ITE step by identifying the all-zero bit string 
$00\cdots0$ of length $n_{\rm steps}$ on the ancillary qubits. 
Then, we evaluate the expectation values of the two terms in Eq.~(\ref{eq:HTFIM}) separately: 
the first term from the $Z$-basis measurements and the second term from the $X$-basis measurements 
of the system qubits.  
We perform 2000 shots for each basis measurement, and the error bars represent the standard 
deviation of the mean.

\begin{figure}[tb]
    \centering
    \includegraphics[width=1.0\columnwidth]{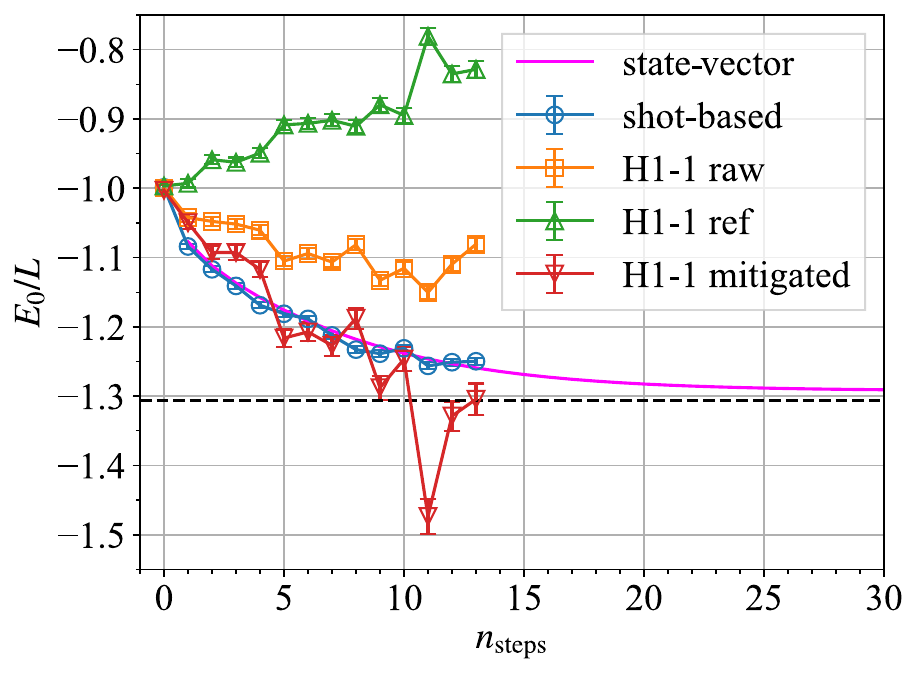}
    \caption{
        Ground-state energy per site $E_0/L$ as a function of the number $n_{\rm step}$ of 
        imaginary-time steps for the TFIM 
        with $L=4$ under PBC, using up to $17$ qubits. 
        The PITE results for the TFIM (orange squares) and those for the reference system 
        (green triangles) are obtained using the H1-1 system with 2000 shots per 
        imaginary-time step without any error mitigation.
        The error-mitigated results, evaluated via Eq.~(\ref{eq:mitigation}), are shown as 
        red inverted triangles.
        For comparison, the results obtained from the noiseless shot-based simulation with 
        the same number of shots (blue circles) and the state-vector simulation 
        (magenta line) are also shown. 
        The dashed horizontal line indicates the exact ground-state energy per site.
    }
    \label{fig:H1-1}
\end{figure}

\subsection{Results and error mitigation}

The energy expectation value without any error mitigation decreases with increasing the imaginary-time steps 
up to approximately $n_{\rm steps}=11$, but then begins to increase for larger $n_{\rm steps}$, 
as shown by the orange squares in Fig.~\ref{fig:H1-1}. 
This behavior contradicts with the state-vector results (magenta line), which show a 
monotonic decrease as a function of $n_{\rm steps}$. 
Furthermore, the initial decrease in energy during the early imaginary-time steps is less pronounced than that 
observed in the state-vector simulations. 
These discrepancies highlight the necessity of applying error mitigation techniques to obtain 
better agreements with the exact state-vector results.

For error mitigation, we assume a global depolarizing noise model on the system qubits 
at each imaginary-time step,  
$\hat{\rho}_{\rm noisy} =  f \hat{\rho}_{\rm ideal} + (1-f) \hat{I}/{2}^{L}$, where 
$\hat{\rho}_{\rm noisy}$ is the density matrix of the system qubits obtained from the noisy PITE 
circuit, 
$\hat{\rho}_{\rm ideal}$ is the corresponding ideal (noise-free) density matrix, 
$\hat{I}/2^{L}$ represents the maximally mixed state of the system qubits, and 
$f$ is an unknown circuit fidelity parameter. 
Under this assumption, the noisy expectation value of the Hamiltonian can be written as 
\begin{equation}
\langle \hat{\cal H}_{\rm TFIM} \rangle_{\rm noisy}  =  f\langle \hat{\cal H}_{\rm TFIM} \rangle_{\rm ideal},
\label{eq:noisyE}
\end{equation}
where $\langle \hat{\cal H}_{\rm TFIM} \rangle_{\rm noisy} = {\rm Tr}[\hat{\cal H}_{\rm TFIM} \hat{\rho}_{\rm noisy}]$ corresponds to the experimentally measured energy expectation value 
(orange squares in Fig.~\ref{fig:H1-1}) and 
$\langle \hat{\cal H}_{\rm TFIM} \rangle_{\rm ideal} = {\rm Tr}[\hat{\cal H}_{\rm TFIM} \hat{\rho}_{\rm ideal}]$ denotes the results from a noiseless shot-based or state-vector simulations 
(blue circles or magenta line in Fig.~\ref{fig:H1-1}).  
In deriving Eq.~(\ref{eq:noisyE}), we used the fact that the Hamiltonian $\hat{\cal{H}}_{\rm TFMI}$ 
is traceless. 
It should be noted that estimating the parameter $f$ is generally difficult, as it requires 
knowledge of $\langle \hat{\cal H}_{\rm TFIM} \rangle_{\rm ideal}$, which is precisely a quantity 
to be evaluated.

To approximate the circuit fidelity parameter $f$, we conduct reference experiments in which the 
real-time evolution operators $\hat{U}_{\rm RTE}(s_1\Delta \tau)$ and $\hat{U}_{\rm RTE}(s_1\Delta \tau)^\dag$ are replaced by alternative reference operators, defined as  
\begin{equation}
\hat{U}_{\rm ref}(s_1\Delta \tau)= 
{\rm e}^{ {\rm i}s_1\Delta \tau \sum_{j}\hat{X}_j}
{\rm e}^{ {\rm i}s_1\Delta \tau \sum_{j}\hat{X}_j \hat{X}_{j+1}}. 
\label{eq:Uref}
\end{equation}
The operator $\hat{U}_{\rm ref}(s_1\Delta \tau)^\dag$ is similarly defined. 
Note that $\hat{U}_{\rm ref}$ represents the first-order Suzuki-Trotter decomposition of the 
time-evolution operator generated by the reference Hamiltonian 
$\hat{\cal{H}}_{\rm ref}=-\sum_{j=1}^L \hat{X}_j\hat{X}_{j+1}-\sum_{j=1}^L\hat{X}_j$, 
which is diagonal in the Pauli-$X$ basis. 
Analogous to Eq.~(\ref{eq:noisyE}), and assuming a global depolarizing noise model in the form $\hat{\rho}_{\rm ref,noisy} =  f_{\rm ref} \hat{\rho}_{\rm ref,ideal} + (1-f_{\rm ref}) \hat{I}/{2}^{L}$, 
we obtain the relation between the noisy and ideal expectation values for the reference experiments 
as 
\begin{equation}
\langle \hat{\cal H}_{\rm TFIM} \rangle_{\rm ref,noisy}  =  f_{\rm ref}\langle \hat{\cal H}_{\rm TFIM} \rangle_{\rm ref,ideal}.
\label{eq:noisyRef}
\end{equation}
Here, $f_{\rm ref}$ denotes the circuit fidelity for the reference experiments, and the other 
quantities are defined in direct analogy with those in the original PITE setup. 
It is important to note that $\hat{U}_{\rm ref}$ is chosen so that the ideal expectation value 
$\langle \hat{\cal H}_{\rm TFIM}\rangle_{\rm ref,ideal}$ is analytically known. 
Specifically, since the initial state $|\psi\rangle = |+\rangle$ is an eigenstate of 
$\hat{U}_{\rm ref}$, the corresponding ideal energy expectation remains constant: 
$\langle \hat{\cal H}_{\rm TFIM}\rangle_{\rm ref,ideal} = -L$, independently of $n_{\rm steps}$.
On the other hand, the experimentally measured noisy expectation value 
$\langle \hat{\cal H}_{\rm TFIM} \rangle_{\rm ref,noisy}$, which typically deviates from the ideal 
value, is directly accessible from the reference experiments (see green triangles in 
Fig.~\ref{fig:H1-1}). 
This allows us to readily estimate the circuit fidelity $f_{\rm ref}$.

We approximate the circuit fidelity $f$ for the original PITE experiments by the corresponding value 
$f_{\rm ref}$ obtained from the reference experiments. 
This leads to the following error-mitigation scheme for estimating the energy expectation value: 
\begin{equation}
    \langle \hat{\cal H}_{\rm TFIM} \rangle_{\rm mitigated}  :=  
    \frac{
    \langle \hat{\cal H}_{\rm TFIM} \rangle_{\rm noisy} 
    }
    {f_{\rm ref}}.
    \label{eq:mitigation}
\end{equation}
A similar error mitigation approach was recently used to study quasi-time-crystalline dynamics of 
local magnetization, as reported in Ref.~\cite{shinjo2024}. 
Although $f_{\rm ref}$ and $f$ may differ in general, we note that the number of ZZPhase gates 
in the compiled circuits for the reference and original experiments is identical at each 
time step. This supports the assumption that 
$f_{\rm ref}$ provides a reasonable approximation for $f$. 
Indeed, the error-mitigated results show better agreement with the ideal values than the 
uncorrected (raw) data, with the exception of the point at $n_{\rm step}=11$ 
(see red inverted triangles in Fig.~\ref{fig:H1-1}). 
This deviation arises from an outlier observed in the reference data at $n_{\rm step}=11$, 
which directly affects the estimated value of $f_{\rm ref}$. 
We were unable to identify the cause of this anomaly. 

Finally, we give three remarks on the error mitigation method used here. 
First, the estimation of $f_{\rm ref}$ based on Eqs.~(\ref{eq:Uref}) and (\ref{eq:noisyRef}) is applicable only when the initial state $|\psi\rangle$ is chosen as the eigenstate of the reference operator $\hat{U}_{\rm ref}$ that has the same number of the two-qubit gates as the original $\hat{U}_{\rm RTE}$. 
Second, since the error mitigation method can violate the variational principle, the lower energy estimate does not necessarily mean the better approximation to the true ground-state energy.  
Indeed, at $n_{\rm steps}=11$, we observe the lowest energy estimate that is far smaller than the exact ground-state energy.
This overshooting of the energy estimate is due to the outlier in the reference data, which underestimates the circuit fidelity $f_{\rm ref}$.
Third, although we have assumed the global depolarizing noise model, there should be errors that cannot be captured by this model. 
For example, the real device has the SPAM error as mentioned above. 
Also, there exist coherent errors such as unwanted single-qubit rotations.
The global depolarizing noise model should be considered as one of the simplest models to mitigate noises.

\section{Summary}

To conclude, we have derived a general description of the PITE algorithm suitable for a 
state-vector simulation. This provides a valuable tool for estimating optimal initial parameters, 
which are strongly dependent on the target Hamiltonian and system size. 
Using these optimal parameters, we demonstrated that the success probability rapidly approaches 
unity after several imaginary-time steps in both Heisenberg and transverse-field Ising models. 
Moreover, by iteratively applying the imaginary-time steps, the energy expectation value reliably converges 
toward the true ground-state energy .

We also performed an experiment on the transverse-field Ising model with $L=4$ sites using 
the trapped-ion quantum computer, Quantinuum H1-1 system. 
The raw experimental results showed limited agreement with those obtained from shot-based 
simulations. 
However, since the number of two-qubit gates in the circuit was at most 355, well below the inverse 
of the two-qubit gate infidelity (approximately 1000), we expected that the fidelity of the raw 
experimental signal relative to the ideal outcome would be at least $0.999^{355} \sim 0.7$ 
for the largest circuit. This 
suggests that meaning signal recovery should be possible through error mitigation. 
To this end, we employed an error mitigation strategy based on a global depolarizing noise model, 
in which the circuit fidelity was approximated by that of a reference circuit. 
After applying this method, the mitigated results showed significantly improved agreement with
the simulation results, except for a single outlier point. 
These results demonstrate that, with appropriately chosen initial parameters and a simple yet 
effective error mitigation scheme, the PITE algorithm can be successfully implemented on current 
quantum hardware, paving the way for its application to larger and more complex quantum systems in the near future on larger quantum computers~\cite{Moses2023}.

\section*{Acknowledgements}

We thank H. Nishi, Y. Nishiya, and Y. Matsushita for enlightening discussions.
This project was made possible by the DLR Quantum Computing Initiative and the Federal Ministry for Economic Aﬀairs and Climate Action; qci.dlr.de/projects/ALQU.
A portion of this work is based on results obtained from Project No. JPNP20017, 
subsidized by the New Energy and Industrial Technology Development Organization (NEDO). 
This study was also supported by JSPS KAKENHI Grants No. JP21H04446 and No. JP22K03520. 
We are further grateful for support from JST COI-NEXT (Grant No. JPMJPF2221) 
and the Program for Promoting Research of the Supercomputer Fugaku (Grant No. MXP1020230411) 
provided by MEXT, Japan.  
Additionally, we acknowledge the support of the UTokyo 
Quantum Initiative, the RIKEN TRIP initiative (RIKEN Quantum), and 
the COE research grant in computational science from Hyogo Prefecture and Kobe City 
through the Foundation for Computational Science.

Note that the PITE algorithm is the subject of an international patent application (international publication number: WO/2023/089930).

\section*{Data availability}
The sample codes for the transverse-field Ising model used in this study are available on Zenodo~\cite{samplecode}. The data presented in this article are also available from the authors upon reasonable request

\appendix
\section*{System-size dependence}\label{appen:size}

\begin{figure}[tb]
 \centering
 \includegraphics[scale=0.8]{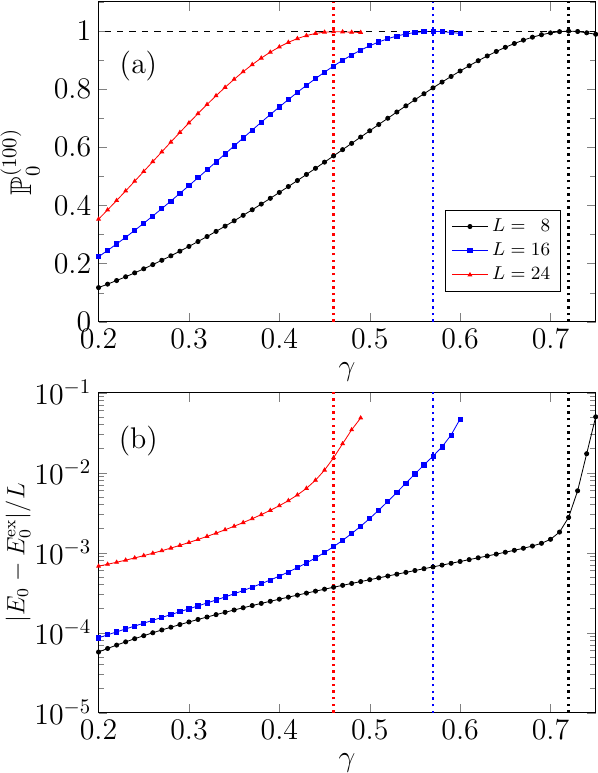}
 \caption{(a) Success probabilities $\mathbb{P}_0^{(100)}$ and (b) the energy difference between the PITE algorithm and those from the ED method after 100 imaginary-time steps with the fixed value of $\Delta\tau=0.2$ for the system sizes $L=8$, $16$, and $24$. The dotted lines indicate $\gamma_{\rm max}$ values for different system sizes $L$.
 }
 \label{fig:P0-Ediff}
\end{figure}

In this appendix, we examine the system-size dependence of the initial parameter $\gamma$ and the resulting energies obtained with the state-vector PITE simulations for system sizes up to $L=32$.

As shown in Fig.~\ref{fig:sv-shot-pite}(a), the success probability $\mathbb{P}_0^{(100)}$ exhibits maxima ($\gamma_{\rm max}$) for fixed $\Delta\tau$ and across different system sizes $L$. This behavior persists for larger system sizes as demonstrated in Fig.~\ref{fig:P0-Ediff}(a), which shows the peak structures around $\gamma_{\rm max}$ (dotted lines). With decreasing $\gamma$, the success probabilities drop rapidly, implying that the number of shots required increases in shot-based simulations or on real devices. 
In general, the energy difference ($\Delta e\equiv |E_0-E_0^{\rm ex}|/L$) between PITE simulations ($E_0/L$) and the ED method ($E_0^{\rm ex}/L$) decreases as $\gamma$ is reduced beyond $\gamma_{\rm max}$, as shown in Fig.~\ref{fig:P0-Ediff}(b). 
While $\Delta e$ is smaller than $10^{-2}$ for the small system size ($L=8$) with $\gamma=\gamma_{\rm max}\simeq0.72$, it exceeds $10^{-2}$ for the larger system sizes $L=16$ and $24$. This is consistent also with the large deviations in energy between the PITE and ED methods for $L=20$ in Fig.~\ref{fig:L20-Heisenberg} with $\gamma=\gamma_{\rm max}(\simeq0.51)$.

\begin{figure}[htb]
 \centering
 \includegraphics[scale=0.8]{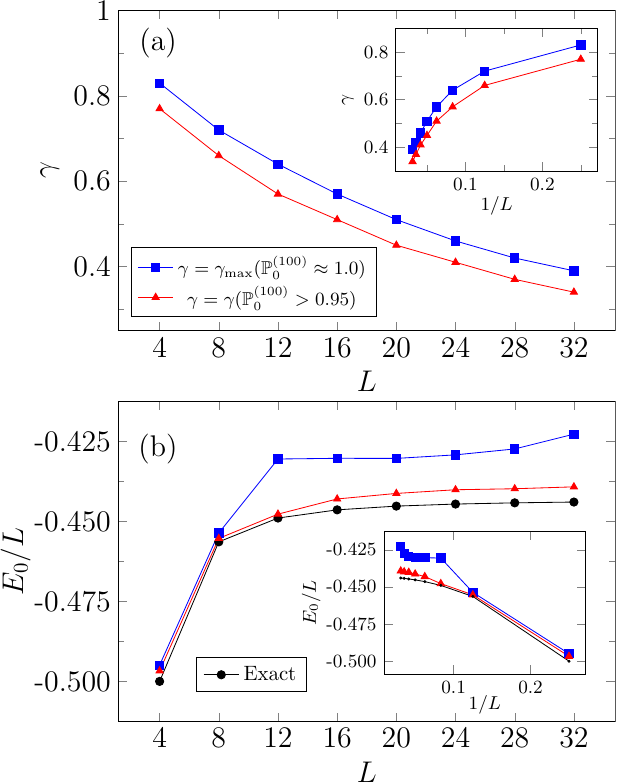}
 \caption{System-size dependence of (a) the initial parameter $\gamma$, which gives us the success probability after 100 imaginary-time steps $\mathbb{P}_0^{(100)}>0.95$ (triangles) and $\mathbb{P}_0^{(100)}\approx1.0$ corresponding to $\gamma_{\rm max}$, and (b) the energy per site $E_0/L$ obtained with these $\gamma$ values for a fixed imaginary-time step $\Delta\tau=0.2$ in the Heisenberg chain given in Eq.~\eqref{eq:Heisen}. 
 The insets of both panels display the $1/L$ dependence. The lines are a guide to the eye.
 }
 \label{fig:system-size-depend}
\end{figure}

Let us now discuss the system-size dependence of $\gamma_{\rm max}$ and the corresponding energy for a fixed $L$. Figure~\ref{fig:system-size-depend}(a) shows the system-size dependence of $\gamma_{\rm max}$ up to $L=32$. As the system size increases, $\gamma_{\rm max}$ decreases systematically. Hence, we expect that finite-size scaling can also be performed for systems with $L>40$, particularly in calculations on large-scale quantum computers that are anticipated in the future.
In Fig.~\ref{fig:system-size-depend}(b), the system-size dependence of the energy using these $\gamma$ values is shown (blue squares) in comparison with the ED results (black circles). For small system sizes ($L\leq8$), the PITE results agree very well with the ED data. On the other hand, for $L\geq12$, deviations become significant as mentioned before. These deviations can be immediately reduced by slightly decreasing $\gamma$. By choosing the smallest $\gamma$ such that the success probability $\mathbb{P}_0^{(100)}\geq0.95$ after 100 imaginary-time steps, the obtained energies [red triangles in Fig.~\ref{fig:system-size-depend}(b)] are very close to the exact ones for larger system sizes. 

Obviously, adaptively tuning $\gamma$ and $\Delta\tau$ during the PITE process represents an important future challenge that could further improve the efficiency and flexibility of PITE-based simulations.

%

\end{document}